\documentclass[aps,prl,twocolumn,superscriptaddress,endnote]{revtex4-1}
\usepackage[latin1]{inputenc}
\usepackage{amsmath}
\usepackage{amsfonts}
\usepackage{amssymb}
\usepackage{graphicx}
\newcommand{\be}{\begin{equation} }
\newcommand{\ee}{\end{equation} }
\newcommand{\ba}{\begin{eqnarray} }
\newcommand{\ea}{\end{eqnarray} }
\newcommand{\n}{\nonumber \\ }

\newcommand{\bv}{\left( \begin{array} }
\newcommand{\ev}{\end{array} \right ) }

\begin{document}
\include{epsfx}

\title{Phase transitions in topological lattice models via topological symmetry breaking}

\author{F. J. Burnell}
\affiliation{All Souls College, Oxford, United Kingdom}
\author{\!\!\!$^{,2}$\,\,\, Steven H. Simon}
\affiliation{Theoretical Physics, Oxford University, 1 Keble Road, Oxford, OX1 3NP, United Kingdom}
\author{J. K. Slingerland}
\affiliation{Department of Mathematical Physics, National University of Ireland, Maynooth, Ireland}
\affiliation{Dublin Institute for Advanced Studies, School of Theoretical Physics,
10 Burlington Rd, Dublin, Ireland}

\begin{abstract}
We study transitions between phases of matter with topological order.  By studying these transitions in exactly solvable lattice models we show how universality classes may be identified and critical properties described.  As a familiar example to elucidate our results concretely, we describe in detail a transition between a fully gapped achiral 2D $p$-wave superconductor ($p+ip$ for pseudospin up/$p-ip$ for pseudospin down) to an $s$-wave superconductor which we show to be in the 2D transverse field Ising universality class.
 \end{abstract}

\maketitle

Motivated in part by growing interest in topological quantum computation, a considerable effort has been invested in understanding systems which realize topologically ordered phases of matter. Though much is understood about the topological properties of these phases and their possible applications to quantum computing\cite{NayakReview,KitaevToric,KitaevVeryLongPaper}, little is presently known about transitions between proximate phases of different topological order.  While a powerful formalism\cite{TSB}, known as ``topological symmetry breaking" (TSB), has been worked out to identify when pairs of phases can in principle be connected by a continuous transition, the critical properties of these transitions have not been systematically studied.   Recently increasing interest has been focused on such phase transitions, highlighting both their importance  to understanding physical systems such as quantum Hall bilayers\cite{WenBaskelli}, and their potential as theoretical models displaying critical behavior unlike that of more conventional statistical mechanical systems\cite{Gilsetal}.

In this paper we take a step towards building a systematic understanding of the relationship between TSB transitions and second-order phase transitions involving a broken symmetry.  Our main result is to describe a large class of  transitions between distinct, topologically non-trivial phases, and show that the effective theory of these transitions is exactly that of the 2D transverse field Ising model (TFIM).   Specifically, we study such phase transitions in the lattice models of topological matter introduced by Levin and Wen\cite{LW}.  We show that for models described by an $SU(2)_k \times \overline{SU(2)}_k$ topological field theory, we can add a perturbation to the Hamiltonian which condenses a bosonic field, and that at energy scales below the quasi-particle gap we can map the perturbed system exactly to a transverse field Ising model.   This  mapping reveals the precise relationship between the phase transition in the topological system and a related {\em global} symmetry-breaking transition.  The lattice models we construct also give interesting proof-of-principle examples of the TSB scenario\cite{TSB}, in which we can track explicitly the fate of the topological order after condensation. 

While the lattice models we study are best viewed as toy models,  the long-wavelength behavior of these toy models should be  applicable in analogous real physical systems.  Consequently we will center our discussion around the example probably most familiar from the literature, where the topological order is that of a
 chiral $p$-wave superconductor\cite{ReadGreen}.  Specifically, we study a Levin-Wen model describing a bi-layer net
achiral  $p$-wave superconductor, in which the order parameter is
$p+ip$ for pseudospin-up (top layer), and $p-ip$ for pseudospin down
(bottom layer). The Ising phase transition we describe in this case is to a phase which is topologically an $s$-wave superconductor.

%One advantage of studying transitions in Levin-Wen models is that these can describe an extremely general set of time-reversal invariant topological phases, allowing us to easily generalize our key results.  The $p+ip/p-ip$ superconductor that we take as an example has topological properties described by an $SU(2)_2 \times \overline{SU(2)}_2$ Chern-Simons theory\footnote{The expert will recognize the difference between $SU(2)_2$, Ising, and other closely related theories.   This difference is not crucial for our discussion.}.  As we will see below, for any $SU(2)_k \times \overline{SU(2)}_k$ theory, there will be an analogous topological phase transition -- and though the number and statistics of the quasi-particles depend on $k$, the phase transition is always that of the TFIM.  In a future work we will describe in detail similar TSB transitions which are related to transitions in the transverse field Potts model\footnote{In this case, for $Q>2$, the phase transitions are known to be first order\cite{TFPotts}}. 

Let us first describe briefly the exactly solvable Levin-Wen lattice Hamiltonians for the honeycomb lattice\cite{LW}.  For simplicity, we focus on the lattice model realizing an  $SU(2)_2 \times \overline{ SU(2)}_2$ topological order, which is that of the achiral $p+ip$/$p-ip$ superconductor.  In this case, each edge in the lattice model can be in one of three states, which we label $1, \sigma$, and $\psi$.  Two types of operators act on these states:  string operators $\hat{s}_{r}$, which raise and lower all edge labels along the trajectory of the string $s$, and vertex projectors $B_V$.  In both cases, their action is determined by the fusion rules of $SU(2)_2$ Chern-Simons theory ($\psi \times \psi=1$, $\sigma \times \psi = \sigma$,  and $\sigma \times \sigma = 1 + \psi$) which describes the topological excitations in the chiral $p$-wave superconductor.  $B_V$ projects onto states in which the three labels entering the vertex $V$ appear in one of the following `allowed' combinations:
%\be(1,1,1) \ \ \ (1, \sigma, \sigma) \ \ \ (1, \psi, \psi) \ \ \ (\sigma, \sigma, \psi)
%\ee
%\begin{center}
%\begin{floatingfigure}[h!]
\be
\includegraphics[height=.5in]{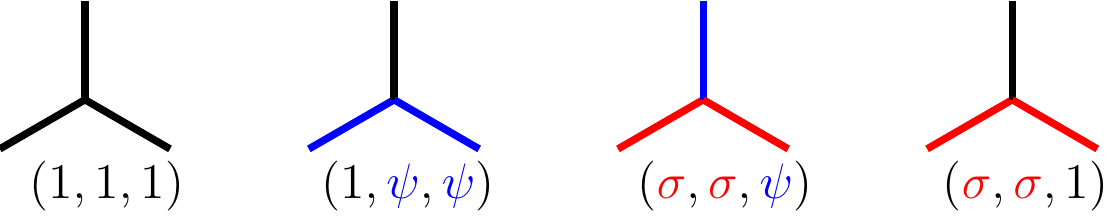}
\ee
%\end{floatingfigure}
%\end{center}
which are chosen such that the three labels incident at each vertex can fuse to give the identity in $SU(2)_2$.  The  string operators act on edge labels according to
\ba \label{Eq_Sops}
\hat{s}_\sigma | 1  \rangle & \sim |\sigma  \rangle  \ \ \   \hat{s}_\sigma | \psi  \rangle  & \sim |\sigma  \rangle  \ \ \    \hat{s}_\sigma | \sigma \rangle  \sim|1  \rangle + |\psi \rangle  \n
\hat{s}_\psi | 1  \rangle &\sim |\psi  \rangle  \ \ \  \hat{s}_\psi | \sigma  \rangle & \sim |\sigma \rangle  \ \ \  \hat{s}_\psi | \psi  \rangle  \sim |1  \rangle
\ea
Here the coefficients of proportionality are dictated by properties of $SU(2)_2$ theory, as described by Ref.~\onlinecite{LW}, but will not be crucial for our discussion.

The Hamiltonian is
\be \label{Eq_Ham}
H = - \sum_V(1- B_V) - \sum_P(1-  B^{(0)}_P)
\ee
where $B^{(0)}_P =\frac{1}{2} \left( 1+\sqrt{2} \hat{s}_\sigma + \hat{s}_\psi \right )$ is a projector composed of string operators acting on the edges bordering the plaquette $P$.  Since all of the projectors in (\ref{Eq_Ham}) commute\cite{LW}, the spectrum of $H$ can be determined exactly.
%The ground state sector has degeneracy $9^{g}$ which depends on the genus $g$ of the surface.
Excitations consist of $8$ types of quasi-particles $\sigma_{L, R}, \, \psi_{L,R}, \, \sigma_L \sigma_R, \, \sigma_L \psi_R, \, \psi_L \sigma_R,$ and $\psi_L \psi_R$.  The subscripts $R$ and $L$ denote the chiralities of the particles.

We may understand the spectrum of this system by considering the corresponding excitations in the superconducting system. The $p+ip$/$p-ip$ superconductor should be thought of as two  independent layers with opposite $(L,R)$ chiralities.   $\sigma_{R,L}$ corresponds to the superconducting vortex (together with its associated $0$-energy Majorana fermion bound state) in the top ($R$) or bottom ($L$) layer; $\psi_{R,L}$ are fermionic quasiparticles in the two layers, and the other $6$ excitations are composites of fermions and/or vortices in both layers.
 The mutual statistics of particles in the same layer are those of the chiral $p$-wave superconductor whereas excitations in different layers have trivial mutual statistics.  Note also that the $\psi$ particles have $\mathbb{Z}_2$ symmetry, meaning they fuse with themselves to give 1.  In the superconducting picture this reflects the fact that fermions are conserved only mod 2.  
%Unlike an $s$-wave superconductor, in the $p$-wave case under consideration $\psi$ quasiparticles may in principle occur either as BdG quasi-particles or as pairs of Majorana zero-mode fermions bound to vortex-cores. 

 In this model,  the particle $\phi \equiv \psi_L \psi_R$ is a boson and can therefore in principle condense in a second-order TSB-type phase transition.    Here we study this transition, showing that as $\phi$ is a $\mathbb{Z}_2$ field, the transition is in the TFIM class.

To condense $\phi$ in the lattice model, we modify the Hamiltonian (\ref{Eq_Ham}) in two ways.   First, noting that $\phi$ violates only plaquette terms\cite{CH}, we  decrease the gap to creating $\phi$ by modifying the plaquette projector:
\be \label{Eq_P1}
B^{(J)}_P = \frac{1}{2} \left( B_P^{(0)} + B_P^{(\phi)} \right) + \frac{J_z}{2} \left( B_P^{(0)} - B_P^{(\phi)} \right)
\ee
where $B^{(\phi)}_P = \frac{1}{2} \left( 1- \sqrt{2} \hat{s}_\sigma + \hat{s}_\psi \right )$ is a projector with eigenvalue $1$ if the plaquette $P$ contains a $\phi$ particle, and $0$ otherwise.
 Second, we add a term
 \be \label{Eq_H1}
H_1 = - J_x \sum_e (-1)^{n_\sigma(e) }
%H_1 = J_x \sum_e n_\sigma(e)  + \mbox{constant}
 \ee
where $n_\sigma(e)$ is 1 if edge $e$ carries label $\sigma$ and is 0 otherwise.   It is easy to check that the operator $(-1)^{n_\sigma(e)}$ applied to the ground state creates $\phi$ particles on the pair of plaquettes bordering $e$ (or hops an existing $\phi$ from one plaquette to another).  $H_1$ is equivalent to adding an energy penalty to all edges carrying the $\sigma$ label.  In the superconducting analogue, the creation of a $\phi = \psi_L \psi_R$ on two neighboring plaquettes is quite a natural term to add to the Hamiltonian  -- corresponding to a four-fermion interaction term that creates or annihilates cooper pairs in the pseudospin (inter-layer)singlet channel.

To describe the resulting phase transition, we exploit the fact that both $H_1$ and $B^{(J)}_P$ commute with $B_V$ at every vertex, and  neither creates plaquette violations other than $\phi$.   We can then study the phase transition in the reduced Hilbert space consisting only of the ground state plus some number of $\phi$ particles created.    Since $\phi$ is a  $\mathbb{Z}_2$ field, we describe the states in the low-energy sector of the lattice model by an Ising variable $n_{\phi} \equiv \frac{1}{2} \left(S_z +1 \right) = 0,1$   on each plaquette (together with a label $\alpha$ identifying  the ground state sector if the system is on the torus).  The vertex projectors, as well as the plaquette term $\frac{1}{2} \left( B_P^{(0)} + B_P^{(\phi)} \right)$, act as the identity on this space.
Within this low-energy sector, the plaquette term $\frac{J_z}{2} \left( B_P^{(0)} - B_P^{(\phi)} \right) $ acts like the spin operator $J_z S_z$, diagonal in the $n_\phi$ basis, while $H_1$ acts like $-J_x \sum_{<ij>} S_x^{(i)} S_x^{(j)}$ as it flips pairs of spins on neighboring plaquettes.  One can check that the operators in Eqs. (\ref{Eq_P1}, \ref{Eq_H1}) satisfy the  appropriate commutation relations.

We thus arrive at an effective theory for the phase transition which is exactly the transverse field Ising model, with spins on sites of the dual lattice.  The paramagnetic phase, in which $J_z$ dominates, corresponds to the initial $SU(2)_k \times \overline{SU(2)}_k$ topological phase; the ferromagnetic phase, in which $J_x$ dominates, contains an indefinite number of $\phi$ particles in each plaquette, and corresponds to the condensed phase.  The transition between these two phases is simply the quantum phase transition of the 2D TFIM.

At this juncture, the reader may wonder how it is that we have mapped a transition which changes the topological order onto one which breaks a global symmetry.  Since topological order is associated with long-ranged statistical interactions (or generalized Berry's phases) which can only occur in the presence of gauge fields, this is indeed a surprising result.  One way to understand it is to note that $\phi$ is essentially the vortex of a $\mathbb{Z}_2$ gauge theory; the special form of the Hamiltonian we use here ensures that no objects behaving like electric sources of this gauge field are present in the ground state anywhere in the phase diagram.  It is because of this that we may exploit the duality between the TFIM and Ising gauge theory\cite{IsingDual} to study the phase transition.  For a more generic choice of Hamiltonian this mapping will no longer hold, as the appropriate dual theory would again be an Ising gauge theory.  However, it is known matter sources do not affect criticality in the Ising gauge theory\cite{VidalToric} -- hence we expect that the transition will still be described by an Ising critical point away from the solvable limit considered here.

Having identified a mechanism for condensing the particle $\phi$, we now describe the topological order of the condensed phase.  We will see that the condensed phase of the lattice model can be mapped exactly onto Kitaev's Toric code (TC)\cite{KitaevToric}, which is known to be the same topological order as that of an $s$-wave superconductor\cite{SondhiSC}.

To understand the topological order of the condensed phase, it is helpful to consider the limit $J_z = 0$, in which the lattice Hamiltonian is again exactly solvable.  Here the plaquette projector takes on a particularly simple form: $B_P =1 + \hat{s}_\psi$, which leaves $n_\sigma$ on each edge unchanged.  When $J_x >0$, this implies that the exact ground state contains no edges labeled $\sigma$, as $H_1$ induces an energy cost $2 J_x$ for each of these.  In this limit,  the label $\sigma$ is confined and we drop it from the low-energy theory.  This leaves a solvable lattice model containing only the two edge labels $1$ and $\psi$, with the Hamiltonian of the Toric code (TC)\cite{KitaevToric} on the honeycomb lattice.

Our next task is to understand how the $8$ excitations of the initial model are related to the TC's $3$.  To identify the spectrum of the condensed phase, we must account for three effects\cite{TSB}.
First, $\psi_L$ and $\psi_R$ are mixed by fusion with the condensate, and become indistinguishable.  Such identification of particles is a generic feature of condensation transitions.  In the language of the superconductor, once a $\langle c_\uparrow c_\downarrow \rangle$ becomes nonzero, fermion number is not conserved mod 2 in each layer independently, but only in the overall system as a whole (i.e., $\mathbb{Z}_2 \times \mathbb{Z}_2$ is broken to a single $\mathbb{Z}_2$).    Second, the label $\sigma$ becomes confined, effectively eliminating the four quasi-particles $\sigma_L, \sigma_R, \psi_L \sigma_R$, and $\sigma_L \psi_R$ from the spectrum.  In the superconducting picture, this is because a vortex in a single layer  engenders an energetically costly `branch cut' in the $s$-wave condensate.
%This is the $SU(2)_2 \times \overline{SU(2)}_2$ system's analogue of the Meissner effect -- we will see that isolated $\sigma_{L,R}$ excitations  engender an energetically costly `branch cut' in the condensate wave function, much like un-quantized magnetic flux in a superconductor.
 Finally, TSB predicts that the bound state $\sigma_L \sigma_R$  {\it splits}  into two distinct types of excitation in the condensed phase\cite{TSB}.
We will show that in the lattice model this splitting can be understood as a result of a new conservation law.  

The energetics of confinement are easily understood in terms of the Ising model representation.   The operator $S^z$, which measures the flux of $\phi$ through a plaquette $P$, is simply $\frac{1}{2} \left( B^{(0)}_P - B^{(\phi)}_P\right ) = \sqrt{2}  \hat{s}_\sigma(P) $ -- the $\sigma$-labeled raising operator applied to all edges surrounding this plaquette.   In the  absence of vertex violations, the  $\sigma$ label always forms closed loops on the lattice, which  constitute domain walls in the $S^x$ basis, since we identify $(-1)^{n_\sigma(ij)} \equiv S^x(i) S^x(j)$.  In the ferromagnetic phase of the Ising model, these domain walls are confined.  The energy cost of a string of edges labeled $\sigma $ is thus linear in the string length, and  the label $\sigma$ disappears from the long-wavelength theory.

Alternatively, we may exploit the duality between the TFIM and Ising gauge theory, given by
\be
S^{z}_P = \pm 1 \rightarrow \prod_{i \mbox{ borders } P} \sigma^z_i = \pm 1 \ \ \ \ \ \ \ \
S^x_i S^x_j = \pm 1 \rightarrow \sigma^z_{\tilde{i}}
\ee
where $\tilde{i}$ is the edge between the two plaquettes $i$ and $j$.  This maps the domain walls of the Ising model -- given in our lattice model by edges where $n_\sigma =1$ -- onto the electric field $\sigma_z =-1$ of the Ising gauge theory.  The symmetry-breaking transition in the TFIM is thus a confining transition in the gauge theory\cite{FradkinShenker}.  This allows us to conclude that open $\sigma$ strings (which join pairs of $\sigma_{L,R}$ vortex excitations) are also confined.
Readers might object that this identification of $\sigma$ strings with the Ising electric flux is not exact, as $\sigma_{L,R}$ are not bosons.  This disagreement in statistics is, however, immaterial to the energetics of confinement, for which we require only {\it static} sources.

\begin{figure}[h!]
\begin{center}
%\vspace*{180pt}
\includegraphics[width=220pt]{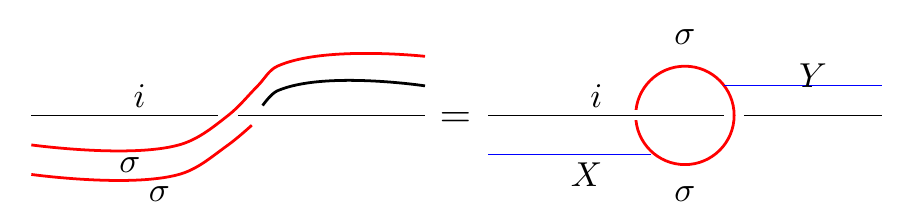}
\caption{ \label{PhaseFig}  The action of the operator $\hat{s}_{\sigma\sigma}$ which creates the quasi-particle $\sigma_L \sigma_R$ which splits in the condensed phase.  The product $\hat{s}_{\sigma_L} \hat{s}_{\sigma_R}$ can be resolved in terms of strings $X, Y = 1, \psi$ which act only on the edge labels, and a phase (depicted diagramatically by the ring labeled $\sigma$) each time the string crosses an edge in the lattice.  (On edges without crossings, the phase must be $1$).  In the uncondensed phase $X,Y = 1$ or $\psi$ on each edge; after condensation $X=Y$ and there are two distinct excitations  $X = Y = \psi$ and $X = Y =1$.
%  The identity shown in (b) ensures that we need only consider diagrams of the type shown in (a).  
}
\end{center}
\end{figure}

To trace the origins of the splitting, we first must understand the form of the operators which create $\sigma_L \sigma_R$.    In the $p+ip/p-ip$ superconductor, this is a pair of vortices, one in each layer, which are bound together.  This excitation may appear in one of two states: the total fermion number of the pair of vortices may be $1$ or $\psi$.  However, this fermion number will change if a {\it single} $\sigma_{L,R}$ vortex in one of the layers braids with this bound pair, and is not a topologically protected quantity.  Hence it is conserved in the phase with $s$-wave pairing, where $\sigma_{L,R}$ vortices are confined, but not in the pure $p$-wave phase.

 In the lattice model, the same effect is seen in the quasi-particle creation operators.  The operator $\hat{s}_{\sigma_{R}} \hat{s}_{\sigma_{L}}$ which creates $\sigma_{L} \sigma_R$ particles acts with a combination of a phase factor at each edge crossed by the string operator, and by acting with $\hat{s}_{1, \psi} \in \{ \hat{s}_1, \hat{s}_\psi \}$ on each edge that $s$ runs along.   (This is depicted diagramatically in Figure \ref{PhaseFig}).  The label $1, \psi$ carried by this string corresponds to the net fermion number of the vortex pair discussed in the previous paragraph.  Table \ref{PhaseTab} gives the action of the components of $\hat{s}_{\sigma \sigma}$.   Here again we find that $\hat{s}_{\sigma \sigma}$ occurs in two distinct flavors, $\hat{s}_{\sigma \sigma}^1$ and $\hat{s}_{\sigma\sigma}^ \psi$, distinguished by whether or not they interchange the edge labels $1$ and $\psi$.
Before condensation only the symmetric combination of these comprises a topological excitation (other combinations have an energy cost which grows with string length).  After confining the $\sigma$ strings, however, we find two distinct quasi-particles, $\sigma \sigma_1$ and $\sigma \sigma_\psi$.  

\begin{table}[h!]
\begin{center}
\begin{tabular}{ccc}
\begin{tabular}{|ccccc| }
\hline
$X$ & $Y$ & $i$ & $i'$ & $\theta$ \\
\hline
$1$ & $1$ & $1$ & $1$ & $0$ \\
$\psi$ & $\psi$ & $1$ & $\psi$ & $0$ \\
\hline
\end{tabular}
&
\begin{tabular}{|ccccc| }
\hline
$X$ & $Y$ & $i$ & $i'$ & $\theta$ \\
\hline
$1$ & $1$ & $\psi$ & $\psi$ & $\pi$ \\
$\psi$ & $\psi$ & $\psi$ & $1$ & $0$ \\
\hline
\end{tabular}
&
\begin{tabular}{|ccccc| }
\hline
$X$ & $Y$ & $i$ & $i'$ & $\theta$ \\
\hline
$\psi$ & $1$ & $\sigma$ & $\sigma$ & $-\pi/4$ \\
$1$ & $\psi$ & $\sigma$ & $\sigma$ & $\pi/4$ \\
\hline
\end{tabular} \\
\end{tabular}
\caption{ \label{PhaseTab} The action of $\hat{s}_{\sigma \sigma}$ on an edge label $i$ has two components: the action of a string  $\hat{s}_X$ which maps $i$ to $i'$, and a phase $\theta$. (Combinations not shown give $0$.)  $X$ and $Y$ may be distinct only if the edge $i$ crossed by the string operator carries the label $\sigma$. After condensation $\sigma$ is confined, and the value of $X$ is conserved along the string.}
\end{center}
\end{table}

%\begin{center}
%\begin{table}
%\begin{tabular}{|c|c|c|c|c|}
%\hline
%$i$ & $(1,1)$ & $(1, \psi)$ & $(\psi, 1 )$ & $(\psi, \psi)$ \\
%\hline
%$1$ & $\frac{1}{\sqrt{2}} 1$ & $0$ & $0$ & $\frac{1}{\sqrt{2}} \psi$ \\
%$\psi$  & $-\frac{1}{\sqrt{2}} \psi $& $0$ & $0$ & $\frac{1}{\sqrt{2}} 1 $ \\
%$\sigma$ &  $0$& $\frac{1}{\sqrt{2}}e^{ \frac{- i \pi}{4} }  \sigma$  & $\frac{1}{\sqrt{2}}e^{ \frac{ i \pi}{4} }  \sigma$&  $0$  \\
%\hline
%\end{tabular}
%\caption{ \label{StringTab} \label{PhaseTab} The action of $\hat{s}_{\sigma \sigma}$ on an edge label $i$ has two components: the action of a string  $\hat{s}_X$ which maps $i$ to $i'$, and a phase $\alpha$.  $(X,Y)$ may be distinct only if the edge $i$ crossed by the string operator carries the label $\sigma$. Combinations not shown give $0$. }
%\end{table}
%\end{center}

To summarize, in the condensed phase we find three quasi-particles $\psi, \sigma\sigma_1$, and $\sigma\sigma_\psi$.  These correspond exactly to the three excitations in the TC:
the two bosons $\sigma \sigma_1$ and $\sigma \sigma_\psi$ are the vortex $m$ and electric source $e$ respectively; the fermion $\psi$ is the combination $em$.  The phases inherited from the original $\sigma_L \sigma_R$ creation operators ensure that braiding any two different particles around each other incurs a phase of $-1$.

To clarify the fate of the topological order across the phase boundary, we consider the ground state degeneracy on the torus.  In the $SU(2)_2 \times SU(2)_2$ phase there are $9$ ground states $|\Omega_\alpha \rangle$, which can be identified by the flux $\alpha$ through one of the
non-contractible curves on the torus.  These fall into two classes, distinguished by the operators $\hat{L}_i = \prod_{e \mbox{ on }c_i} (-1)^{n_\sigma (e) }$, where $c_i, i=1,2$ run along edges in the dual lattice around the two non-contractible curves on the torus.  Four of the ground states have at least one $L_i=-1$.  Because $\sigma$ can only
appear in closed loops, this means that these states must have a string of edges labeled $\sigma$ which run across the width of the system in one of the two directions; in the confined phase, where such a string incurs an energetic
cost linear in the system size, they are no longer ground states.  Of the remaining five, the two anti-symmetric combinations $(|\Omega_{\psi_L} \rangle- |\Omega_{\psi_R} \rangle$ and $(|\Omega_{\psi_L \psi_R} \rangle- |\Omega_{1} \rangle )$ vanish in the condensed ground state.  Hence these also become gapped and split off from the ground-state sector, leaving only the two symmetric combinations  $(|\Omega_{\psi_L} \rangle+ |\Omega_{\psi_R} \rangle$ and $(|\Omega_{\psi_L \psi_R} \rangle+ |\Omega_{1} \rangle )$.  Finally, as $\psi$ appears only in closed loops after condensation, $|\Omega_{\sigma_L \sigma_R} \rangle$ splits into two states, distinguished by the eigenvalue of $\prod_{e \mbox{ on }c_1} (-1)^{n_\psi (e) }$.  This gives four ground states on the torus, as expected for the TC.

The transition studied so far, between $SU(2)_2 \times \overline{SU(2)}_2$ and the TC, is prototypical of a class of transitions between pairs of topological phases, all of which exhibit transverse-field Ising criticality.   We postpone a detailed description of these other transitions to a future work; here we merely outline their general features:  Consider a Levin-Wen Hamiltonian describing the topological theory $SU(2)_k \times \overline{SU(2)}_k$, which has quasi-particles of spin $0, \frac{1}{2}, ... \frac{k}{2}$ in two mutually non-interacting sectors $R$ and $L$.  We condense the boson $\phi_k \equiv \frac{k}{2}_L \frac{k}{2}_R$ by modifying the plaquette projector as in Eq. (\ref{Eq_P1}), and adding a boson creation term $\sum_e (-1)^{2s(e)}$, where $s(e)$ is the spin on edge $e$.  As $\phi_k$ is always a $\mathbb{Z}_2$ field which commutes with the vertex projectors, the dynamics of the phase transition is described by the transverse field Ising model.
% (along with the requisite label to distinguish the ground states).
%The gap to all non-condensing quasi-particles remains finite throughout the transition.
All half-odd integer spin edge labels map to
the domain wall;
% (the mapping is many to one as the Ising model only faithfully describes one of the ground states on the torus); 
hence particles of net half-odd integer spin (of the form $\frac{2j}{2}_L \frac{2i+1}{2}_R$, or vice versa) are confined in the condensed phase.  As before, pairs of particles $(i_L j_R, \frac{ k-2i}{2}_L \frac{k-2j}{2}_R$ which mix by fusion with $\frac{k}{2}_L \frac{k}{2}_R$ are no longer distinct in the condensed phase.  This leaves $\frac{k^2}{4} + \frac{k}{2} +1$ potential quasi-particles.

As in the $k=2$ case, it remains to ask whether any of these quasi-particles will split.    If $k$ is odd, there is no splitting.
This leads to a theory whose topological structure is $SO(3)_k \times \overline{SO(3)}_k$ -- meaning that the only effect of condensation is to eliminate half-odd integer spin excitations from the theory.   The final theory is simple in this case because the initial particle spectrum is of the form $\{ a_1, a_2, ... a_{r/2} ; \phi \times a_1, \phi \times a_2, ... \phi \times a_{r/2} \}$, and condensation merely confines particles of net half-odd integer spin, and identifies pairwise excitations $\frac{2j+1}{2}_L \frac{2i+1}{2}_R \equiv \frac{ k - 2j -1}{2}_L \frac{k-2j-1}{2}_R$.
If $k$ is even, the criteria of Ref. \onlinecite{TSB} dictate that the $\frac{k}{4}_L \frac{k}{4}_R$ particle splits into two components.  In the lattice model, these are distinguished by whether the associated edge string operator carries even or odd integer spin.  In this case the final theory is not a doubled theory with decoupled right- and left- handed sectors, but a more involved type of achiral theory known as a Drinfeld double\cite{Drinfeld}.

In fact, the protocol outlined here can be extended to a wider class of transitions for models where an achiral $\mathbb{Z}_q$ boson is condensed, resulting in a transition of the $q$-state transverse field Potts type.   
%We will discuss the structure of these more general theories in a forthcoming work.

%More generally, consider a lattice model whose excitations take the form $C \times \overline{C}$, where $C$ contains an  excitation $\psi$ such that $\psi^k =1$.  One may condense the achiral boson $\phi \equiv \psi_L \times \psi_R$, which violates only plaquettes and obeys $\phi^k =1$.  The effective theory of the phase transition in this case is the transverse field Potts model.  We defer a more detailed description of these models to a future work.

Here we have described a phase transition between phases exhibiting the topological orders of a $p+ip$/$p-ip$ superconductor, and that of an $s$-wave superconductor (or equivalently, the Toric code\cite{KitaevToric,SondhiSC}).  By realizing the initial phase in a solvable Levin-Wen\cite{LW} type lattice model, where the phase transition can be induced by a simple deformation of the Hamiltonian, we show that  and that the phase transition is that of the $2$D transverse-field Ising model.   The analogous transitions for the topological order of $SU(2)_k \times \overline{SU(2)}_k$ can also be studied in the lattice model, leading to a family of condensation transitions which are all in the same Ising universality class.  

{\bf Acknowledgements:} SHS acknowledges funding from an SFI ETS Walton fellowship. JS is supported by SFI PI Award 08/IN.1/I1961.  The authors are grateful to S.L. Sondhi, M.A. Levin, and B. Halperin for helpful discussions, and for the hospitality of the Aspen Center for Physics.
%, Station Q, and KITP. 

\bibliography{TSBBib}

\end{document}